\documentclass[reqno]{amsart}

\newtheorem{lemma}{Lemma}
\newtheorem{theorem}{Theorem}

\newtheorem{hypothesis}{Hypothesis}
\newtheorem{remark}{Remark}

\newcommand{\be}{\begin{eqnarray}}
\newcommand{\ee}{\end{eqnarray}}
\newcommand{\bee}{\begin{eqnarray*}}
\newcommand{\eee}{\end{eqnarray*}}
\newcommand{\R}{\mbox {\sc R}}

\newcommand{\C}{\mbox {\sc C}}

\newcommand{\I}{\mbox {\sc 1}}

\newcommand{\Rp}{\mbox {\sc r}}

\newcommand{\asy}{{\it O}}
\newcommand{\ind}{\hskip 0.5cm}

\begin{document}

\title [nonlinear Schr\"odinger equation]{Nonlinear time-dependent one-dimensional Schr\"odinger equation with double well potential}

\author {Andrea Sacchetti}

\address {Dipartimento di Matematica\\
Universit\`a di Modena e Reggio Emilia\\
Via Campi 213/B, I--41100 Modena, Italy}

\date {\today}

\email {Sacchetti@unimo.it}

\thanks {This work is partially supported by the Italian MURST and INDAM-GNFM. \ I thank Prof. Vincenzo Grecchi and Prof. Andr\'e Martinez for helpful discussions and remarks.}

    \begin {abstract}
We consider time-dependent Schr\"odinger equations in one dimension with double well potential and an external nonlinear perturbation. \ If the initial state belongs to the eigenspace spanned by the eigenvectors associated to the two lowest eigenvalues then, in the semiclassical limit, we show that the reduction of the time-dependent equation to a $2$-mode equation gives the dominant term of the solution with a precise estimate of the error. \ By means of this stability result we are able to prove the destruction of the beating motion for large enough nonlinearity.
     \end{abstract}


\maketitle

\section {Introduction}

\ind Recently, the theoretical analysis of the nonlinear time-dependent Schr\"odinger equation
\be
i \hbar \dot \psi = H_0 \psi + \epsilon |\psi |^2 \psi , \ \ \epsilon \in \R ,\ \dot \psi = \frac {\partial \psi }{\partial t}, \label {equa1}
\ee
where
\bee
H_0 = - \frac {\hbar^2}{2m} \Delta + V , \ \  \ \Delta = \sum_{j=1}^d \frac {\partial^2}{\partial x_j^2}, \ d \ge 1 ,
\eee
has attracted an increasing interest (see \cite {SS} for a review and \cite {LSY} for a rigorous derivation of the Gross-Pitaevskii energy functional). \ When $V$ is a double well potential, one of the main goal is to understand how the nonlinear perturbation with strength $\epsilon$ affects the unperturbed beating motion (see, e.g., the review paper \cite {DGPS} and the paper \cite {V} where equation (\ref {equa1}) is proposed as a model for chiral molecules). \ To this end, it is crucial to study the solution $\psi$ for times of the order of the beating period; in other words, for practical purposes the unit of time is given by the beating period $T=\pi \hbar /\omega$ where $\hbar$ is the Planck's constant and $\omega$ is one half of the energy splitting between the two lowest energies.

\ind Here, I consider equation (\ref {equa1}) in the semiclassical limit where, by assuming that $d=1$ and under some generic assumption on the double well potential, we give the asymptotic behavior of the solution $\psi$ with a precise estimate of the error. \ In particular, the main result (Theorem 3) consists in proving that the solution of the Gross-Pitaevskii equation is approximated, with a rigorous control of the error, by means of the solution of a two-dimensional dynamical system exactly solvable. \ As a result it follows (Theorem 4) that the beating motion between the two wells of a state initially prepared on the two lowest eigenstates gradually disappears for increasing nonlinearity.

\ind A similar investigation has been recently performed in \cite {GMS}, where the nonlinear perturbation is given by $\epsilon \langle \psi , g \psi \rangle g \psi$ and $g(x)$ is a given odd function, and in \cite {S}, where, in dimension $d=1$ and $d=3$, we consider the limit of large barrier between the two wells. \ In particular, in \cite {S} I had to assume that the discrete spectrum of the Schr\"odinger operator $H_0$ consists of only two non-degenerate eigenvalues and that the restriction to the continuous eigenspace of the unitary evolution operator satisfies to a priori estimate uniformly with respect to the parameters of the model.

\ind Finally, we mention other recent results concerning the study of the existence of stationary solutions for Gross-Pitaevskii equations with double well potentials \cite {AFGST}, \cite {AMP} and, in the case of single-well type potentials, the existence of solutions asymptotically given by solitary wave functions in the case that the discrete spectrum of the linear Schr\"odinger operator has only one non-degenerate eigenvalue \cite {SW}, \cite {W}.

\ind Our paper is organized as follows. 

\ind In Section 2 we introduce the main notations and we state the assumptions on the potential. \ Moreover, we collect some semiclassical results concerning the spectrum of the linear Schr\"odinger operator.

\ind In Section 3 we prove the global existence of the solution of the Gross-Pitaevskii equation, the existence of conservation laws and a priori estimate (Theorem 2). \ The global existence of the solution is proved for both repulsive and attractive nonlinear perturbation, where, in the second case, we have to assume that the strength of the nonlinear perturbation is small enough.

\ind In Section 4 we introduce the two-level approximation which, roughly speaking, consists in projecting the Gross-Pitaevskii equation onto the two-dimensional space spanned by the eigenvectors of the linear Schr\"odinger operator associated to the two lowest eigenvalues. \ For practical purposes, it is more convenient to choose, as a basis of such a two-dimensional space, the two "single-well" states. \ The dynamical system we obtain is exactly solvable.

\ind In Section 5 we give our main result (Theorem 3) proving the stability of the two-level approximation. \ Here, we make use of the comparison criterion between ordinary differential equations and of a priori estimate of the solution of the Gross-Pitaevskii equation. \ We underline that, in order to obtain such an estimate, the assumption $d=1$ on the dimension plays a crucial role.

\ind In Section 6 we give the full rigorous justification of the results by Vardi \cite {V} proving the existence of a critical value for the nonlinearity parameter giving the destruction of the beating motion (Theorem 4).

\section{Assumptions and preliminary results} 

\ind Here, we consider the Cauchy problem
\be
i \hbar \dot \psi = H_{\epsilon } \psi , \ \ H_{\epsilon } = H_{0} + W \label {equa2}
\ee
\bee
\psi (0,x) = \psi^0 (x) \in L^2 (\R ), \ \ \| \psi^0 \| =1,   
\eee
where $\dot \psi$ denotes the derivative of $\psi$ with respect to the time $t$, $H_0$ is the linear Schr\"odinger operator formally given by (here, $x$ denotes the spatial variable in dimension 1)
\be
H_{0} = - \frac {\hbar^2}{2 m} \frac {d^2}{dx^2} + V , \label {equa3}
\ee
$V$ is a symmetric double well potential  and 
\bee
W = \epsilon |\psi |^2 , 
\eee
is the nonlinear perturbation with strength $\epsilon$.

\ind In the following, for the sake of definiteness, we denote by $C$ any positive constant independent of $\epsilon$, $\hbar$ and $t$, we assume $\hbar$ small enough, that is $\hbar \in (0, \hbar^\star ]$ for some $\hbar^\star$, and we denote 
\bee
\| \varphi \|_p = \| \varphi \|_{L^p} = \left \{ \int |\varphi (x)|^p d x \right \}^{1/p} \ \mbox { and } \ \| \varphi \| = \| \varphi \|_2 .
\eee
Moreover, given $y = (y_1 ,\ldots , \ y_m ) \in \R^m$ for some $m \ge 1$, we denote 
\be
|y| = \max_{1\le j \le m}|y_j |. \label {equa4}
\ee 

\subsection {Assumptions on the potential} \ Here, we assume that the potential $V$ is a regular symmetric function which admits two non-degenerate minima and it is bounded from below. \ More precisely:

\begin {hypothesis} 
{
The potential $V (x)$ is a real valued function such that:

\begin {itemize}

\item [{\it i.}] $V(-x)=V(x)$, $\forall x \in \R$;

\item [{\it ii.}] $V \in C^2 (\R )$;

\item [{\it iii.}] $V (x)$ admits two non-degenerate minima at $x=\pm a$, for some $a>0$, such that 
\be
V (x) > V_{min} = V(\pm a) , \ \ \forall x \in \R , \ x\not= \pm a ; \label {equa5}
\ee
in particular, for the sake of definiteness, we assume that 
\bee
\frac { d V (\pm a )}{d x} =0 \ \ \mbox { and } \ \ \frac {d^2 V (\pm a )}{dx^2} >0 ; 
\eee

\item [{\it iv.}] finally we assume that 
\bee
{\liminf}_{|x|\to \infty}  V(x) =V_\infty >V_{min}.
\eee

\end {itemize}
}
\end {hypothesis}

\ind It follows that the operator formally defined in (\ref {equa3})  admits a self-adjoint realization (still denoted by $H_0$) on $L^2 (\R )$ (see, for instance, Theorem III.1.1 in \cite {BS}). \ 
 Let $\sigma (H_0 ) = \sigma_d \cup \sigma_{ess}$ be the spectrum of the self-adjoint operator $H_0$ where $\sigma_d$ denotes the discrete spectrum and $\sigma_{ess}$ denotes the essential spectrum. \ From Hyp. 1-{\it iv} it follows that $\sigma_d \subset (V_{min},V_\infty )$, $\sigma_{ess} = \emptyset $ if $V_\infty = + \infty$ (see Theorem XIII.67 in \cite {RS4}) and that $\sigma_{ess} \subseteq [V_\infty ,+\infty )$ if $V_\infty < \infty $ (see Theorem III.3.1 in \cite {BS}). \ Furthermore, the following two Lemmas hold:

\begin {lemma}
Let $\sigma_d$ be the discrete spectrum of $H_0$. \ Then, for any $\hbar \in (0,\hbar^\star ]$ it follows that:

\begin {itemize}

\item [{\it i.}] $\sigma_d$ is not empty and, in particular, it contains two eigenvalues at least;

\item [{\it ii.}] let $\lambda_{1,2}$ be the lowest two eigenvalues of $H_0$, they are non-degenerate, in particular $\lambda_1 < \lambda_2$ and there exists $C>0$, independent of $\hbar$, such that 
\bee
\inf_{\lambda \in \sigma (H_0) - \{ \lambda_{1,2} \} } [\lambda - \lambda_{2} ] \ge C \hbar . 
\eee

\end {itemize}

\end {lemma}

\begin {proof}
The proof is an immediate consequence of the above assumptions and standard WKB arguments.
\end {proof}

\begin {lemma} 
Let $\varphi_{1 ,2}$ be the normalized eigenvectors associated to $\lambda_{1,2}$, then:

\begin {itemize}

\item [{\it i.}] $\varphi_{j}$, $j=1,2$, can be chosen to be real-valued functions such that $\varphi_{j} (-x) = (-1)^{j-1} \varphi_{j} (x)$;

\item [{\it ii.}] $\varphi_{j } \in H^1 (\R )$;

\item  [{\it iii.}] $\varphi_j \in L^p (\R )$ for any $p \in [1,+\infty ]$;

\item [{\it iv.}] there exists a positive constant $C$ such that
\be
\| \varphi_{j } \|_p \le C \hbar^{- \frac {p-2}{4 p}} , \ \ \forall p \in [2,+\infty ], \ \ \forall \hbar \in (0, \hbar^\star ]. \label {equa6}
\ee
\end {itemize}
\end {lemma}

\begin {proof}
Property {\it i.} immediately follows from assumption Hyp. 1-{\it i}. \ Property {\it ii.} follows from Lemma III.3.1 in \cite {BS}. \ Property {\it iii.} follows from Theorem III.3.2 in \cite {BS}. \ Finally, property {\it iv.} follows for $p=+\infty $ by means of standard WKB arguments. \ From this fact, from the normalization of the eigenvectors  and from the H\"older inequality then property {\it iv.} follows for any $p\in [2,+\infty ]$:
\bee
\| \varphi_j \|_p = \left [ \| \varphi_j^2 \varphi_j^{p-2} \|_1 \right ]^{1/p} \le \| \varphi_j \|_2^{2/p} \| \varphi_j \|_\infty^{(p-2)/p}=\| \varphi_j \|_\infty^{(p-2)/p}.
\eee
\end {proof}

\subsection {Splitting and single-well states}  \ind It is well know that the splitting between the two lowest eigenvalues vanishes as $\hbar $ goes to zero. \ In particular, we have that:

\begin {lemma} 
Let 
\bee
\omega = \frac {\lambda_{2}-\lambda_{1}}{2} \ \mbox { and }\ \Omega = \frac {\lambda_{2}+\lambda_{1}}{2}
\eee
and
\bee
\varphi_{R} = \frac {1}{\sqrt 2} \left [ \varphi_{1} + \varphi_{2} \right ] \ \mbox { and } \ \varphi_{L} = \frac {1}{\sqrt 2} \left [ \varphi_{1} - \varphi_{2} \right ] 
\eee
where $\varphi_{1 ,2}$ are the normalized eigenvectors associated to $\lambda_{1,2}$. \ Then there exist two positive constants $C$ and $\Gamma$, independent of $\hbar$, such that 
\be
\| \varphi_{R } \varphi_{L} \|_\infty \le C \omega 
\label {equa7}
\ee
and
\be
\omega \le C e^{- \Gamma /\hbar } , \ \forall \hbar \in (0,\hbar^\star ]. \label {equa8}
\ee
As a result if follows that  
\be
\lim_{\hbar \to 0} \omega = 0 \label {equa9}
\ee
and 
\be
\lim_{\hbar \to 0} \frac {\Omega - V_{min}}{\hbar } =c  \label {equa10}
\ee
for some $c>0$.
\end {lemma}

\begin {proof} In order to prove this Lemma we observe that $V$ is a symmetric double well potential with non-zero barrier between the wells. \ That is, let $\delta >0$ be small enough and let us define the two sets 
\bee
\left. \begin {array} {r} 
B_{R } = \left \{ x \in \R^+ \ : \ V (x) \le V_{min} + \delta  \right \} \\ 
B_{L} = \left \{ x \in \R^- \ : \ V (x) \le V_{min} + \delta \right \} 
\end {array} 
\right \} \ \mbox { i.e. } \ \ x \in B_{R} \Leftrightarrow -x \in B_{L}
.
\eee
From condition (\ref {equa5}) it follows also that 
\bee
B_{R } =[b,c] \ \mbox { and } \  B_{L } =[-c,-b] 
\eee
for some $c>a>b>0$. \ The sets $B_{R,L}$ are usually called "wells". \ Let
\bee
\Gamma_\delta = \int_{-b}^b \sqrt {\max [V (x)-(V_{min} + \delta ), 0]} d x >0,
\eee
be the Agmon distance between the two wells. \ From these facts and from standard WKB arguments  (see \cite {H} and \cite {HS}) then (\ref {equa7})---(\ref {equa10}) follow for some $\Gamma \in [\Gamma_0 , \Gamma_\delta ]$.
\end {proof}

\begin {remark}
{\rm 
By definition it follows that $\varphi_{R } (-x)=\varphi_{L}(x)$; moreover, from (\ref {equa7}), it follows that these functions are localized on only one of the "wells" $B_{R}$ and $B_{L}$, e.g.:
\bee
\int_{B_{R}} |\varphi_{R} (x)|^2 dx = 1 + \asy (e^{-C /\hbar })
\eee
for some $C>0$. \ For such a reason we call them "single-well" (normalized) states.
}
\end {remark}

\begin {remark}
{\rm 
We underline that, by assuming some regularity properties on the potential $V$, then it is possible to obtain the precise asymptotic behavior of the splitting as $\hbar $ goes to zero \cite {HS}.
}
\end {remark}

\subsection {Assumptions on the parameter} \ We assume that the two parameters $\omega$ and $\epsilon$ are such that 
\bee
\omega \to 0 \ \ \mbox { and } \ \ \epsilon \to 0 \ \ \mbox { as } \ \ \hbar \to 0 
\eee
and there exists a positive constant $C$ such that
\be
\frac {c \epsilon }{\omega } \le C , \ \ c=\| \varphi_R^2 \| , \ \ \forall \hbar \in (0,\hbar^\star ] . \label {equa11}
\ee

\subsection {Assumption on the initial state} \ Let 
\be
\Pi_c = \I - \langle \varphi_{R } , \cdot \rangle \varphi_{R } - \langle \varphi_{L} , \cdot \rangle \varphi_{L} \label {equa12}
\ee
be the projection operator onto the eigenspace orthogonal to the 2-dimensional eigenspace associated to the doublet $\{ \lambda_{1,2} \}$. \ Let $\psi^0$ be the initial wavefunction, we assume that:

\begin {hypothesis}
$\Pi_c \psi^0 =0$.
\end {hypothesis}

\section {Global existence of the solution and conservation laws}

\ind Here, we prove that the Cauchy problem (\ref {equa2}) admits a solution for all time provided that assumptions Hyp.1--2 are satisfied and the strength $\epsilon$ of the nonlinear perturbation is small enough. \ Moreover, we prove a "priori" estimate of the solution $\psi$. 

\ind The following results hold.

\begin {theorem} There exist $\hbar^\star >0$ and $\epsilon_0>0$ such that for any $\hbar \in (0,\hbar^\star ]$ and $\epsilon \in [-\epsilon_0 , \epsilon_0 ]$ then the Cauchy problem (\ref {equa2}) admits a unique solution $\psi (t,x) \in H^1$ for any $t \in \R$. \ Moreover, the following conservation laws hold:
\be
\| \psi (t,\cdot )\| = \| \psi^0 (\cdot )\| = 1 \label {equa13}
\ee
and
\be
E (\psi ) = \frac {\hbar^2}{2m} \left \| \frac {\partial \psi}{\partial x} \right \|^2 + \langle V \psi , \psi \rangle + \frac 12 \epsilon \| \psi^2 \|^2 = E (\psi^0 ) . \label {equa14}
\ee
\end {theorem}

\begin {proof}
From Hyp.2 it follows that 
\bee
\psi^0 = c_1 \varphi_{1 }+c_2 \varphi_{2 }, \ \ c_{1,2} = \langle \psi^0 , \varphi_{1,2 }\rangle.
\eee
From this fact and from Lemma 2 then $\psi^0 \in H^1$. \ Therefore, existence of the global solution $\psi \in C(\R ,H^1)$ and the conservation laws (\ref {equa13}) and (\ref {equa14}) follow from known results (see, e.g., the papers quoted in \cite {SS} and \cite {SW}) for any $\epsilon >0$ (repulsive nonlinear perturbation) and for any $\epsilon \in (-\epsilon_0 ,0)$ for some $\epsilon_0 >0$ (attractive nonlinear perturbation).  
\end {proof}

\begin {remark} 
{\rm 
There exists a positive constant $C$ independent of $\hbar$ and $\epsilon$ such that 
\be
|E (\psi)- V_{min}|\le C ( \omega + \hbar + \epsilon \hbar^{-1/2 }), \ \ \forall \hbar \in (0,\hbar^\star  ], \ \ \forall \epsilon \in [- \epsilon_0 , \epsilon_0] .\label {equa15}
\ee
This estimate immediately follows from (\ref {equa14}), from the assumption Hyp.2 and from Lemmas 1 and 2. \ Indeed, from Hyp.2 it follows that
\bee
E(\psi^0 ) = \left \langle H_0 (c_1 \varphi_1 + c_2 \varphi_2 ), (c_1 \varphi_1 + c_2 \varphi_2 ) \right \rangle + \frac 12 \epsilon \| \psi^0 \|_4^4 
\eee
where $\| \psi^0 \|_4 \le C \hbar^{-1/8}$ from (\ref {equa6}) and where 
\bee
\left \langle H_0 (c_1 \varphi_1 + c_2 \varphi_2 ), (c_1 \varphi_1 + c_2 \varphi_2 ) \right \rangle &=& \lambda_1 |c_1|^2 + \lambda_2 |c_2|^2 = \Omega - \omega + 2 \omega |c_2|^2.
\eee
From these facts and from (\ref {equa10}), then (\ref {equa15}) follows.
}
\end {remark}

\begin {theorem}
Let $\epsilon_0 (\hbar )$ be a function such that 
\be
\lim_{\hbar \to 0} \epsilon_0 (\hbar ) /\hbar^2 =0. \label {equa16}
\ee
The solution $\psi$ of equation (\ref {equa2}) satisfies to the following uniform estimate: there exists a positive constant $C$ independent of $t$, $\hbar$ and $\epsilon$, such that 
\be
\| \psi \|_p \le C \left [ \frac {|E(\psi^0)-V_{min}| }{\hbar^2}\right ]^{\frac {p-2}{4 p}}, \ \ \forall p \in [2,+\infty ],   \label {equa17}
\ee
and
\bee
\left \| \frac {\partial \psi}{\partial x } \right \| \le C \left [ \frac {|E(\psi^0) - V_{min}|}{\hbar^2}\right ]^{\frac 12}
\eee
for all time and $\forall \hbar \in (0, \hbar^\star ]$, $\forall \epsilon \in [-\epsilon_0(\hbar ) , \epsilon_0 (\hbar )]$.
\end {theorem}

\begin {proof}
In order to prove the estimate (\ref {equa17}) let 
\bee
 k= \frac {\hbar}{\sqrt {2m}}, \ \Lambda = \frac {E(\psi^0 ) - V_{min}}{k^2} .
\eee
Then, the conservation laws (\ref {equa13}) and (\ref {equa14}) imply that 
\bee
\left \| \frac {\partial \psi}{\partial x } \right \|^2 + \frac 12 [\mbox {sign}(\epsilon )] \rho^2 \| \psi^2 \|^2  \le \Lambda ,
\eee
where 
\bee 
\rho = |\epsilon |^{1/2}  /k \ll 1, 
\eee
since (\ref {equa16}). \ In particular, if we set 
\bee
\chi = \rho \psi  
\eee
then the above equation takes the form 
\bee
\left \| \frac {\partial \chi}{\partial x } \right \|^2 + \frac 12 [\mbox {sign}(\epsilon )]\| \chi^2 \|^2  \le \Lambda \rho^2
\eee
from which it follows that 
\be
\left \| \frac {\partial \chi}{\partial x } \right \|^2 \le \rho^2 |\Lambda | + \frac 12 \| \chi^2 \|^2 = \rho^2 |\Lambda |+ \frac 12 \| \chi \|_4^4 \label {equa18}
\ee
From the Gagliardo-Niremberg inequality (see, for instance, \cite {FIP} and \cite {Wi}, where the dimension is here equal to 1)
\be
\| \chi \|_{2\sigma +2}^{2\sigma +2} \le C \left \| \frac {\partial \chi}{\partial x } \right \|^{\sigma } \| \chi \|^{2+\sigma }, \ \ \forall \sigma \ge 0 ,
\label {equa19}
\ee
where we choose $\sigma =1$, it follows that
\bee
\| \chi \|_4^4 \le C \left \| \frac {\partial \chi}{\partial x } \right \| \| \chi \|^3 \le C \left \| \frac {\partial \chi}{\partial x } \right \| \rho^3
\eee
since $\| \chi \| = \rho \| \psi \| = \rho $ and $\| \psi \| =1$. \ By inserting this inequality in (\ref {equa18}) it follows that $\left \| \frac {\partial \chi}{\partial x } \right \|$ satisfies to 
\be
\left \| \frac {\partial \chi}{\partial x } \right \|^2 \le \rho^2 |\Lambda | + C  \rho^{3} \left \| \frac {\partial \chi}{\partial x } \right \| 
\label {equa20}
\ee
for any $t\in \R$. \ From (\ref {equa20}) immediately follows that 
\bee
\left \| \frac {\partial \chi}{\partial x } \right \| \le \sqrt {|\Lambda |} \rho \left ( 1+ o(1) \right ) ,\ \ \mbox { as }\ \rho \to 0 . 
\eee
Hence, $\left \| \frac {\partial \psi}{\partial x } \right \| \le C \sqrt {|\Lambda |}$ and, from (\ref {equa19}), we have that 
\bee
\| \psi \|_p \le C \left \| \frac {\partial \psi}{\partial x } \right \|^{\sigma/p} \le C |\Lambda |^{(p-2)/4p} 
\eee
where we choose now $\sigma = \frac {p-2}{2}$, i.e. $p=2\sigma +2$.
\end {proof}

\begin {remark}
{\rm 
Condition (\ref {equa16}) is true in the semiclassical limit and under the assumption (\ref {equa11}). 
}
\end {remark}

\begin {remark}
{\rm 
From the fact $E(\psi_0 ) -V_{min}=\asy (\hbar )$, which follows from (\ref {equa8}), (\ref {equa15}) and (\ref {equa16}), and from the bounds (\ref {equa17}) and (\ref {equa11}) then it follows that $\forall t \in \R$, $\forall \hbar \in (0, \hbar^\star ]$, $\forall \epsilon \in [-\epsilon_0(\hbar ) , \epsilon_0 (\hbar )]$ then 
\be
\| \psi \|_p \le C \hbar^{-\frac {p-2}{4 p}} , \ \forall p \in [2, +\infty ],\ \ \mbox { and } \ \ \left \| \frac {\partial \psi}{\partial x } \right \| \le C \hbar^{-\frac 12}  . 
\label {equa21}
\ee
}
\end {remark}

\section{Two-level approximation}

\ind For our purposes it is more convenient to make the substitution $\psi \to e^{-i\Omega t /\hbar} \psi $, hence equation (\ref {equa2}) takes the following form (where, with abuse of notation, we still denote the new function by $\psi$)
\be
i \hbar \dot \psi = (H_0 - \Omega ) \psi + \epsilon |\psi |^2 \psi , \ \ \psi (x,0)= \psi^0 (x). \label {equa22}
\ee
\ind Let us write the solution of this equation in the form 
\be
\psi (t,x) = a_R (t) \varphi_{R}(x) + a_L (t) \varphi_{L} (x) + \psi_c (t,x) , \label {equa23}
\ee
where $a_R(t)$ and $a_L(t)$ are unknown complex-valued functions depending on time and $\psi_c = \Pi_c \psi$, $\Pi_c $, defined in (\ref {equa12}), is the projection onto the space orthogonal to the two-dimensional space spanned by the two "single-well" states $\varphi_{R}$ and $\varphi_{L}$, i.e.:
\bee
\langle \psi_c , \varphi_{R} \rangle = \langle \psi_c , \varphi_{L} \rangle = 0 , \ \ \forall t \in \R .
\eee
From the conservation law (\ref {equa13}) it follows that
\be
|a_R (t)|^2 + |a_L (t)|^2 + \| \psi_c (t, \cdot ) \|^2 = 1 ,\forall t \in \R .\label {equa24}
\ee

\ind By substituting $\psi$ by (\ref {equa23}) in equation (\ref {equa2}) we obtain that $a_R$, $a_L$ and $\psi_c$ must satisfy to the system of differential equations
\be
\left \{ 
\begin {array}{lcl}
i \hbar \dot a_R &=&  - \omega a_L +  \epsilon \langle \varphi_{R} , |\psi |^2 \psi \rangle \\ 
i \hbar \dot a_L &=&  - \omega a_R +  \epsilon \langle \varphi_{L} , |\psi |^2 \psi \rangle \\ 
i \hbar \dot \psi_c &=& (H_0 - \Omega ) \psi_c + \epsilon \Pi_c |\psi |^2 \psi 
\end {array}
\right. \label {equa25}
\ee

\ind By substituting again $\psi$ by (\ref {equa23}) in the first two equations of the above system then we obtain that these equations take the form 
\be
\left \{ 
\begin {array}{lcl}
i \hbar \dot a_R &=& - \omega a_L + \epsilon c |a_R|^2 a_R + \epsilon r_R  \\ 
i \hbar \dot a_L &=& - \omega a_R + \epsilon c |a_L|^2 a_L + \epsilon r_L  
\end {array}
\right. \label {equa26}
\ee
where  
\be
c= \| \varphi_{R}^2 \|^2 = \| \varphi_{L}^2 \|^2 = \asy (\hbar^{-1}) \label {equa27}
\ee
and where $r_R$ and $r_L$ are given by
\bee
r_R &=&  \langle \varphi_R , |\psi |^2 \psi \rangle - |a_R^2| a_R \langle \varphi_R , |\varphi_R|^2 \varphi_R \rangle \\
&=& \langle \varphi_R , |\psi |^2 \phi_L \rangle + a_R \langle |\varphi_R |^2, |\phi_L |^2 + a_R \varphi_R \bar \phi_L + \bar a_R \bar \varphi_R \phi_L \rangle \\
r_L &=&  \langle \varphi_L , |\psi |^2 \psi \rangle - |a_L^2| a_L \langle \varphi_L , |\varphi_L|^2 \varphi_L \rangle \\
&=& \langle \varphi_L , |\psi |^2 \phi_R \rangle + a_L \langle |\varphi_L |^2, |\phi_R |^2 + a_L \varphi_L \bar \phi_R + \bar a_L \bar \varphi_L \phi_R \rangle 
\eee
where
\bee
\phi_L = a_L \varphi_L + \psi_c \ \ \mbox { and } \ \  \phi_R = a_R \varphi_R + \psi_c 
\eee

\ind We denote {\bf two-level approximation} the solutions $b_R$ and $b_L$ of the system of ordinary differential equations 
\be
\left \{ 
\begin {array}{lcl}
i \hbar \dot b_R &=&  - \omega b_L + \epsilon c |b_R|^2 b_R   \\ 
i \hbar \dot b_L &=&  - \omega b_R + \epsilon c |b_L|^2 b_L 
\end {array}
\right. , \ \  b_{R,L} (0) = a_{R,L} (0).
\label {equa28}
\ee
obtained by neglecting the remainder terms $r_R$ and $r_L$ in (\ref {equa26}). \ It is easy to see that the solution of this system satisfies to the following conservation law
\be
|b_R(t)|^2 + |b_L(t)|^2 = |b_R(0)|^2 + |b_L(0)|^2 = |a_R(0)|^2 + |a_L(0)|^2 =1, \label {equa29}
\ee
and, moreover, it is also possible to explicitely compute (see \cite {RSFS} and Appendix B in \cite {S}) the solution of (\ref {equa28}) by means of elliptic functions. \ In particular, we obtain that the imbalance function, defined as 
\be
z(t) =|b_R(t)|^2 - |b_L (t)|^2, \label {equa30}
\ee
is given by
\bee
z(t ) = 
\left \{ 
\begin {array}{l} 
A \mbox {cn} \left [ A \eta (\omega t/\hbar -\tau_0)/2k ,k \right ] , \ \ \mbox { if } k <1 , \\ 
A \mbox {dn} \left [ A \eta (\omega t /\hbar -\tau_0)/2,1/k \right ] , \ \ \mbox { if } k >1  , \end {array} \right. 
\eee
where $\eta =\epsilon c/\omega$, $\tau_0$ depends on the initial condition, 
\bee
I= \sqrt {1-z^2 (0)} \cos [\theta (0)] - \eta z^2 (0)/4,
\eee
$\theta = \arg (b_R) - \arg (b_L)$ is the relative phase, 
\bee
A = \frac {2 \sqrt 2}{\eta} \left [ \sqrt {\frac 14 \eta^2  +1 + I \eta } - \left ( 1 + \frac 12 I \eta \right ) \right ]^{1/2}, 
\eee
and 
\be
k^2 = \frac {1}{2} \left [ 1 - \frac {1 + \frac 12 I \eta }{\sqrt {\frac 14 \eta^2  +1 + I \eta }} \right ] . \label {equa31}
\ee
We underline that $z(t)$ periodically assumes positive and negative values if, and only if, $k<1$. 

\section{Stability of the two-level approximation}

\ind Our main result consists in proving the stability of the two-level approximation when we restore the remainder terms $r_R$ and $r_L$ in equation (\ref {equa28}). 

\ind We prove that:

\begin {theorem}
Let $\psi_c = \Pi_c \psi$, $a_{R}(t)= \langle \psi , \varphi_{R } \rangle $ and $a_{L}(t)= \langle \psi , \varphi_{L } \rangle $, where $\psi$ is the solution of equation (\ref {equa22}), let $b_R (t)$ and $b_L (t)$ be the solution of the system of ordinary differential equations (\ref {equa28}). \ Let $\epsilon \in [-\epsilon_0 (\hbar ) , \epsilon_0 (\hbar )]$, where $\epsilon_0 (\hbar )$ satisfies the condition (\ref {equa16}). \ Then, for any $\tau ' >0$ there exists a positive constant $C$ independent of $\epsilon $, $\hbar$ and $t$ such that:
\be
\left | b_{R,L} (t) - a_{R,L} (t) \right | \le C e^{-C \hbar^{-1}} 
 \ \mbox { and } \ \| \psi_c (\cdot , t ) \| \le C e^{-C \hbar^{-1}} 
\label {equa32}
\ee
for any $\hbar \in (0,\hbar^\star ]$ and for any $t\in [0,\hbar \tau' /\omega ]$. 
\end {theorem}

\begin {proof} For the sake of simplicity, hereafter, let us drop out the parameters where this does not cause misunderstanding. \ In order to prove the theorem we introduce the "slow time" $\tau = \omega t/\hbar $ and let
\bee
\left \{ 
\begin {array}{l}
A_{R,L}(\tau ) = a_{R,L}(t)  \\ \ B_{R,L}(\tau )=b_{R,L}(t)
\end {array}
\right. , \ \ R_{R,L}(\tau ) = \frac {\epsilon}{\omega} r_{R,L}(t) \ \mbox { and } \ \eta = \frac {\epsilon c}{\omega} . 
\eee
Then (\ref {equa26}) and (\ref {equa28}) respectively take the form (here $'$ denotes the derivative with respect to $\tau$)
\be
\left \{ 
\begin {array}{lcl}
A_R' &=& i A_L -i  \eta |A_R|^2 A_R +  R_R  \\ 
A_L' &=& i A_R - i \eta |A_L|^2 A_L +  R_L   
\end {array}
\right. \label {equa33}
\ee
and 
\be
\left \{ 
\begin {array}{lcl}
B_R' &=& i B_L -i  \eta |B_R|^2 B_R \\ 
B_L' &=& i B_R - i \eta |B_L|^2 B_L    
\end {array}
\right. \label {equa34}
\ee
satisfying to the same initial condition
\bee
B_{R,L} (0) = A_{R,L} (0) = a_{R,L}(0), 
\eee
since (\ref {equa24}) and (\ref {equa29}), they are such that 
\be
|B_R (\tau )|^2 + |B_L (\tau )|^2 =1 , \ \ |A_R (\tau )|^2 + |A_L (\tau )|^2 \le 1. \label {equa35}
\ee
In a more concise way, with an obvious meaning of the notation, we can write (\ref {equa33}) and (\ref {equa34}) as
\be
A' = f(A) + R \ \ \mbox { and }\ \ B' = f(B), \ \ A(0)=B(0) = a(0), \label {equa36}
\ee
where $A,B \in S^2$ since (\ref {equa35}), $S^2 = \{ (z_1,z_2) \in \C^2 : |z_1|^2+|z_2|^2 \le 1\}$.

\begin {lemma}
The function $f: S^2 \to C^2$ satisfies to the Lipschitz condition: 
\be
\left | f(A) - f(B) \right | \le L |A-B|, \ L=1+3\eta . \label {equa37}
\ee
\end {lemma}

\begin {proof} According with the notation (\ref {equa4}) we have that 
\bee
|f(A)-f(B)| = \max \left [ |f_R|,|f_L| \right ]
\eee
where $|A|\le 1$ and $|B|\le 1$ since $A,\ B \in S^2$ and where 
\bee
f_R &=& (A_L - B_L)- \eta (|A_R|^2 A_R -|B_R|^2 B_R ) \\ 
f_L &=& (A_R - B_R)- \eta (|A_L|^2 A_L -|B_L|^2 B_L )
\eee
Then (\ref {equa37}) immediately follows since  
\bee
f_R = (A_L-B_L) - \eta \left [ |B_R|^2 (A_R-B_R) + A_R (|A_R|+|B_R|)(|A_R|-|B_R|) \right ]
\eee
where $\left | |A_R|-|B_R| \right | \le |A_R-B_R|$ and where the other term $f_L$ will be treated in the same way. 
\end {proof}

\begin {lemma}
Let  
\bee
\beta = \max [c \epsilon , \omega ] 
\eee
where $c$ is defined in (\ref {equa27}). \ Let $\psi_c = \Pi_c \psi$ where $\psi$ is the solution of equation (\ref {equa22}); it satisfies to the following uniform estimate
\be
\| \psi_c \| \le C \beta \hbar^{-3/2} \left [ \exp [ C \hbar^{-1/2} (\epsilon t/\hbar )] +1 \right ] , \ \ \forall t \in  \R  , \label {equa38}
\ee
for some positive constant $C$ independent of $\hbar ,\ \epsilon $ and $t$. 
\end {lemma}

\begin {proof} As a first step we consider the following raw estimates:
\be
\| \psi_c \|_p \le C \hbar^{- \frac {p-2}{4 p}}, \ \ \forall p \in [2,+\infty ], \ \ \forall t \in \R , \label {equa39}
\ee
and
\bee
|r_{R,L} | \le C \hbar^{-1/2}, \ \ \forall t \in \R . 
\eee
Indeed, (\ref {equa39}) immediately follows from the Minkowski inequality and from (\ref {equa21}):
\bee
C \hbar^{- \frac {p-2}{4 p}} \ge \| \psi \|_p \ge - \left ( |a_R(t)| \| \varphi_R \|_p + |a_L (t)| \| \varphi_L \|_p \right ) + \| \psi_c \|_p 
\eee
where $|a_{R,L} (t) |\le 1$ and where $\varphi_{R,L}$ satisfy the bound (\ref {equa6}). \ In the same way, from Lemma 2 and Theorem 2, it follows that 
\bee
|r_{R}| &\le & C \| \varphi_R \psi^2 \| \cdot \| \psi \| + \| \varphi_R \|_4^4 \\ 
&\le & C \| \varphi_R \|_\infty \| \psi \|^2_4 \| \psi \| + C \| \varphi_R \|_4^4  \\
&\le & C \hbar^{-1/2}
\eee
and similarly for $|r_L|$. 

\ind Now, in order to prove the estimate (\ref {equa38}) we make use of the third equation of (\ref {equa25}) from which it follows that 
\bee
\psi_c (\cdot ,t) = -i \frac {\epsilon }{\hbar }\int_0^t e^{-i (H_0-\Omega) (t-s)/\hbar}\Pi_c |\psi ( \cdot ,s)|^2 \psi ( \cdot ,s) ds  
\eee
since $\psi_c^0 =\Pi_c \psi^0 =0$ from assumption Hyp.2. 

\ind Let $\psi = \varphi + \psi_c$ where $\varphi = a_R \varphi_R + a_L \varphi_L$, then 
\bee
|\psi |^2 \psi = \varphi_I + \psi_c \varphi_{II} + \bar \psi_c \varphi_{III} ,\ 
\left \{ \begin {array}{l} 
\varphi_I = |\varphi |^2 \varphi ,\\ \varphi_{II} = 2 |\varphi |^2 + 2 \bar \psi_c \varphi + |\psi_c |^2 +  \bar \varphi \psi_c ,\\ \varphi_{III} = \varphi^2 
\end {array}
\right.
.
\eee
Therefore, we can write
\bee
\psi_c = -i \frac {\epsilon }{\hbar }\left [ I + II + III \right ]
\eee
where
\bee
I &=& \int_0^t e^{-i (H_0 - \Omega )(t-s) /\hbar } \Pi_c \varphi_I ds \\
II &=& \int_0^t e^{-i (H_0 - \Omega )(t-s) /\hbar } \Pi_c \psi_c \varphi_{II} ds \\
III &=& \int_0^t e^{-i (H_0 - \Omega )(t-s) /\hbar } \Pi_c \bar \psi_c \varphi_{III} ds 
\eee
For what concerns the first term we have that, by integrating by part, 
\bee
I &=& \left [ -i \hbar e^{-i (H_0 - \Omega )(t-s) /\hbar } [H_0 - \Omega ]^{-1} \Pi_c |\varphi |^2 \varphi \right ]_0^t + \\ 
&& \ \ \ + i \hbar \int_0^t e^{-i (H_0 - \Omega )(t-s) /\hbar } [H_o - \Omega ]^{-1} \Pi_c \frac {\partial |\varphi |^2 \varphi }{\partial s} ds
\eee
Let us underline that from Lemma 1 it follows that the following operators, from $L^2$ into $L^2$, are bounded 
\bee
\left \| e^{-i (H_0 - \Omega )(t-s)/\hbar } \right \| =1 ,\ \ \ \| \hbar [H_0 - \Omega ]^{-1} \Pi_c \| \le C ,
\eee
and, from Lemma 2 and equations (\ref {equa24}) and (\ref {equa26}) we have the following uniform estimate for any $t\in \R$ 
\bee 
\| \dot \varphi \|_p &\le &\left ( |\dot a_r | + |\dot a_L|\right ) \left (\| \varphi_R \|_p + \| \varphi_L \|_p \right ) \le C \hbar^{-1} \max [c \epsilon , \omega ,\epsilon \hbar^{-\frac 12} ] \hbar^{- \frac {p-2}{4 p}} \\ &\le & C \hbar^{-1} \beta \hbar^{- \frac {p-2}{4 p}} .
\eee
Then we have that
\bee
\| I \| &\le & C \max_{s\in [0,t]} \left \{ \| \varphi^3 (s,\cdot )\| + t \| \dot \varphi (s,\cdot )\varphi^2 (s,\cdot )\| \right \} \\ 
&\le & C   \max_{s\in [0,t]} 
\left \{ \| \varphi (s,\cdot )\|_6^3 + t \| \dot \varphi (s,\cdot )\|\cdot \| \varphi (s,\cdot )\|^2_\infty \right \} \\ 
&\le & C \left \{ \hbar^{- 1/2} + t \hbar^{-1} \beta \hbar^{-1/2} \right \} .
\eee
For what concerns the other two terms we have that 
\bee
\| II \| \le \int_0^t \| \psi_c \| \cdot \| \varphi_{II} \|_\infty ds \le C \hbar^{-1/2} \int_0^t \| \psi_c \| ds 
\eee
since $\| \varphi_{II} \|_\infty \le C \hbar^{-1/2}$, and similarly 
\bee
\| III \| \le \int_0^t \| \psi_c \| \cdot \| \varphi_{III} \|_\infty ds \le C \hbar^{-1/2} \int_0^t \| \psi_c \| ds .
\eee
Indeed, from Lemma 2 and (\ref {equa39}) it follows that 
\bee
\| \varphi_{II} \|_\infty \le C \left \{ \| \varphi \|_\infty^2 + \| \psi_c \|_\infty \| \varphi \|_\infty + \| \psi_c \|^2_\infty \right \} \le C \hbar^{-1/2}
\eee
and
\bee
\| \varphi_{III}\|_\infty \le \| \varphi \|^2_\infty \le C \hbar^{-1/2}.
\eee
Collecting all these results and denoting
\bee
g(t) = \| \psi_c (\cdot ,t ) \|
\eee
we have that $g(t)$ is a positive real valued function satisfying the estimate
\bee
g(t) &\le & C \frac {\epsilon }{\hbar } \left \{ \hbar^{-1/2} \int_0^t g(s) d s + \hbar^{-1/2} \left ( 1 + t \hbar^{-1} \beta \right ) \right \} \\ 
& \le & a \int_0^t g(s) ds + a+a b t , \ \ a= C \frac {\epsilon}{\hbar^{3/2}} ,\ \ b=  \frac { \beta }{\hbar} .
\eee
From this estimate, since $\psi_c (0) =0$ and from the Gronwall's Lemma (see \cite {Hi}, pag. 19) it follows that
\bee
g(t) &\le & a+a b t + a \int_0^t e^{a (t-s)} (a+a b s) ds =-b+a e^{a t}+be^{at} \\ &\le & \frac {C \beta}{\hbar^{3/2}} \left [e^{C \epsilon t \hbar^{-3/2}} +1 \right ]
\eee
proving so the result.
\end {proof}

\ind From the inequality (\ref {equa8}) and from the assumption (\ref {equa11}) it follows that for any fixed $\tau' >0$ then there exists $C>0$ satisfying the second inequality in (\ref {equa32}).

\begin {lemma}
For any fixed $\tau' >0$ then the remainder terms $r_{R}$ and $r_L$ satisfy to the following uniform estimate:
\bee
\max \left [ |r_R|,|r_L | \right ] \le C \beta \hbar^{-2 } e^{C \hbar^{-1/2 }}, \ \ \forall t \in [0, \tau' \hbar /\omega ] , 
\eee
for some positive constant $C$ independent of $\hbar$, $\epsilon$ and $t$.
\end {lemma}

\begin {proof}
Let us only consider the term $|r_R|$, the other term $|r_L|$ could be treated in the same way. \ By definition and since $\max [ |a_R|,|a_L| ] \le 1$ then it follows that 
\be
|r_R | 
&\le &  + \left | \langle \varphi_{R}  \bar \varphi_{L} , |\psi|^2 \rangle \right | \label {equa40} \\ 
&& \ \ + \left | \langle \varphi_{R}  | \psi |^2 , \psi_c \rangle \right | \label {equa41} \\ 
&& \ \ + \left | \langle |\varphi_R |^2, |\phi_L |^2 + a_R \varphi_R \bar \phi_L + \bar a_R \bar \varphi_R \phi_L \rangle \right | 
\label {equa42} 
\ee 
and we estimate separately each term.

\ind From Lemma 3, equation (\ref {equa13}) and the H\"older inequality it follows that the term (\ref {equa40}) satisfies to the following estimate:
\bee
\left | \langle \varphi_{R} \varphi_{L} , |\psi |^2 \rangle \right | \le \| \varphi_{R} \bar \varphi_{L} \|_\infty \cdot \| \psi^2 \|_1 \le C \omega .
\eee

\ind From Lemma 5 and the H\"older inequality it follows that the term (\ref {equa41}) satisfies to the following estimates:
\bee
\left | \langle \varphi_{R} |\psi |^2 , \psi_c \rangle \right | \le \| \varphi_{R} \|_\infty \cdot \| \psi^2 \| \cdot \| \psi_c \| \le C \beta \hbar^{-2 } e^{C\hbar^{-1/2 }}
\eee
and that the term (\ref {equa42}) satisfies to the estimate
\bee
&& \left | \langle |\varphi_R |^2, |\phi_L |^2 + a_R \varphi_R \bar \phi_L + \bar a_R \bar \varphi_R \phi_L \rangle \right | \le \\ && \ \ \le C \left [ \| \varphi_R \varphi_L \|_\infty + \| \varphi_R^2 \|_\infty \| \psi_c \|^2 +  \| \varphi_R \varphi_L \|_\infty \| \psi_c \| \right ] \le C \omega
\eee
Collecting all these estimates we obtain the proof of the Lemma. 
\end {proof}

\smallskip

\ind The proof of the Theorem is almost done. \ Indeed, equations (\ref {equa36}) can be rewritten in the integral form:
\bee
A(\tau ) = A(0) + \int_0^\tau f[A(s)] d s + \int_0^\tau R d s 
\eee
and
\bee 
B(\tau ) = B(0) + \int_0^\tau f [B(s)] ds 
\eee
from which it follows that for any $\tau \in [0,\tau ']$ 
\bee
|A(\tau )-B(\tau )| &\le & \int_0^\tau \left | f[A(s)] - f[B(s)]\right | ds + \int_0^\tau |R| d s \\
&\le & a \int_0^\tau \left | A(s) - B(s) \right | ds +  b \tau , \ a=L, \ b= C \frac {\epsilon \beta \hbar^{-2} e^{C \hbar^{-1/2}}}{\omega}
\eee
since Lemmas 4 and 5. \ From this inequality and by means of the Gronwall's Lemma we finally obtain that 
\bee
|A(\tau ) - B(\tau )| &\le & b \tau + a b \int_0^\tau e^{a (t-s)} s ds = \frac ba \left [ e^{a \tau } -1 \right ] \nonumber \\
&\le & \frac {C}{L} \frac {\epsilon \beta \hbar^{-2} e^{C \hbar^{-1/2}}}{\omega}
\eee
proving so (\ref {equa32}) since
\bee
\frac {\omega + \epsilon}{C' \omega} \le L=1+3\eta \le C' \frac {\omega + \epsilon }{\omega}, 
\eee
for some $C'>0$, which implies that $\frac {\beta }{L \omega } \le C$ for some $C>0$.
\end {proof}

\begin {remark}
{\rm Recalling that $\omega = \asy ( e^{-\rho /\hbar })$ then the above theorem implies that for any $\alpha <1$ and for any $\tau' >0$ there exists $C$ such that 
\bee
|b_{R,L} (t) - a_{R,L} (t) | \le C \omega^\alpha \ \ \mbox { and } \ \ \| \psi_c (\cdot , t ) \| \le C \omega^\alpha , \ \ \forall t \in [0 , \hbar \tau' /\omega ] .
\eee
}
\end {remark}

\section {Destruction of the beating motion for large nonlinearity}

\subsection {The unperturbed case $\epsilon =0$} \ Under the assumption Hyp.2 it follows that the solution of the unperturbed equation
\bee
i \hbar \dot \psi = H_{0} \psi , \ \ \psi (0,x)=\psi^0 (x),
\eee
is simply given by
\bee
\psi (t,x) &=& e^{-i\Omega t/\hbar } \left [ \frac {c_1 + c_2}{\sqrt{2}} \cos (\omega t/\hbar ) + i \frac {c_2 -c_1}{\sqrt{2}} \sin (\omega t/\hbar ) \right ] \varphi_{R} (x) + \\ 
&& \ \ + e^{-i\Omega t/\hbar} \left [ \frac {c_1-c_2}{\sqrt{2}} \cos (\omega t/\hbar) - i \frac {c_1 + c_2}{\sqrt{2}} \sin (\omega t/\hbar) \right ] \varphi_{L} (x)
\eee
where
\bee
c_{1,2} = \langle \varphi_{1,2},\psi^0 \rangle , \ \ |c_1|^2 + |c_2|^2 =1.
\eee
Hence, $\psi (t,x)$ is, up to the phase factor $e^{-i(\Omega -\omega )t/\hbar}$, a periodic function with period $T=\pi \hbar / \omega$. 

\ind In particular, if $\psi$ initially coincides with a single-well state, e.g. $\psi^0 = \varphi_{R}$, then  
\bee
\psi (t,x) = e^{-i(\Omega - \omega)t/\hbar } \left [ e^{-i\omega t/\hbar} \cos (\omega t/\hbar)  \varphi_{R} (x)- i e^{-i\omega t/\hbar} \sin (\omega t/\hbar) \varphi_{L} (x)\right ]
\eee
and the state $\psi (t,x)$ performs a beating motion. \ That is the state, initially localized on the well $B_{R }$, is localized on the other well $B_{L}$ after half a period and, after a whole period, it "returns" on the initial well, and so on. \ In particular, let us consider the motion of the "center of mass" defined here as  
\bee
\langle X \rangle^t = \langle X \psi , \psi \rangle = \int_{\Rp} X(x) |\psi (t,x)|^2  dx 
\eee
where $X \in C (R) \cap L^2 (\R)$ is a given bounded function such that $X(-x)=-X(x)$. \ We have that 
\bee
\langle X \rangle^t = \left [ \cos^2 (\omega t/\hbar) - \sin^2 (\omega t/\hbar) \right ] \int_{\Rp} X(x) |\varphi_{R} (x)|^2  dx 
\eee
is a periodic function which periodically assumes positive and negative values, i.e. we have the well know beating motion for double-well problem.

\subsection {The perturbed case $\epsilon \not= 0$} \ In such a case it follows that the "center of mass" is given by
\bee
\langle X \rangle^t = c [|a_R(t)|^2- |a_L(t)|^2] + r, \ \ c = \langle X \varphi_R , \varphi_R \rangle ,
\eee
where the remainder term $r$ satisfies the uniform estimate 
\bee
|r| &=& 2 \left | \Re \left [ a_R \bar a_L \langle X \varphi_R , \varphi_L \rangle + \langle X \psi , \psi_c \rangle \right ] \right | \\ 
&\le & 2 \left [ \| \varphi_R \varphi_L \|_\infty + \| X\|_\infty \| \psi \| \| \psi_c \| \right ] \\ 
&\le & C e^{-C \hbar^{-1}} ,\ \ \forall t \in [0, \hbar \tau' /\omega ].
\eee
If we denote by $z(t)$ the imbalance function defined in (\ref {equa30}) then it follows that 
\bee
|a_R(t)|^2- |a_L(t)|^2 =z(t) + C e^{-C \hbar^{-1}},\ \ \forall t \in [0, \hbar \tau' /\omega ],
\eee
hence
\bee
\langle X \rangle^t = c z(t) + C e^{-C \hbar^{-1}},\ \ \forall t \in [0, \hbar \tau' /\omega ].
\eee

\ind Then, we have that:

\begin {theorem}
Let Hyp. 1 and 2 be satisfied. \ Let $k^2$ be defined as in (\ref {equa31}), it depends on the initial wavefunction $\psi^0$. \ Let $\tau' >0$ fixed, $\langle X \rangle^t$ is, up to a remainder term, a periodic function for any $t\in [0,\hbar \tau' /\omega ]$. \ In particular, if:

\begin {itemize}

\item [i)] $k^2<1$ then $\langle X \rangle^t$ periodically assumes positive and negative values (i.e. the beating motion still persists);

\item [ii)] $k^2>1$ then $\langle X \rangle^t$ has a definite sign (i.e. the beating motion is forbidden).

\end {itemize}

\end {theorem}

\begin {remark}
{\rm Let us close by underlining that when the wavefunction is initially prepared on just one well, e.g. $\psi^0 = \varphi_R$, then 
\bee
I= - \frac 14 \eta \ \ \mbox { and } \ \ k^2 = \frac {1}{16} \eta^2.
\eee
Therefore, from the theorem above it follows that for $|\eta |$ larger than the critical value $4$ the beating motion is forbidden. \ In such a way, we put on a full rigorous basis the results obtained by \cite {V} in the two-level approximation.
}
\end {remark}

\end {document}